\input{aipcheck}

\documentclass[
    ,final            % use final for the camera ready runs
  ]
  {aipproc}

\layoutstyle{8x11single}

\begin{document}

\title{Simultaneous Radio to (Sub-) mm-Monitoring of Variability and Spectral Shape Evolution of potential GLAST Blazars}


\author{L. Fuhrmann}{
  address={Max-Planck-Institut f\"ur Radioastronomie (MPIfR), Auf dem H\"ugel 69, 53121 Bonn, Germany}
}

\author{J. A. Zensus}{
  address={Max-Planck-Institut f\"ur Radioastronomie (MPIfR), Auf dem H\"ugel 69, 53121 Bonn, Germany}
}

\author{T. P. Krichbaum}{
  address={Max-Planck-Institut f\"ur Radioastronomie (MPIfR), Auf dem H\"ugel 69, 53121 Bonn, Germany}
}

\author{E. Angelakis}{
  address={Max-Planck-Institut f\"ur Radioastronomie (MPIfR), Auf dem H\"ugel 69, 53121 Bonn, Germany}
}

\author{A. C. S. Readhead}{
  address={California Institut of Technology, 1200 E. California Blvd., Pasadena, CA 91125}
}

\begin{abstract}
The Large Area Telescope (LAT) instrument onboard GLAST offers a tremendous opportunity for 
future blazar studies. In order to fully benefit from its capabilities and to maximize the 
scientific return from the LAT, it is of great importance to conduct dedicated multi-frequency 
monitoring campaigns that will result comprehensive observations. Consequently, we initiated 
an effort to conduct a GLAST-dedicated, quasi-simultaneous, broad-band flux-density (and 
polarization) monitoring of potential GLAST blazars with the Effelsberg and OVRO radio 
telescopes (11\,cm to 7\,mm wavelength). Here, we present a short overview of these 
activities which will complement the multi-wavelengths activities of the 
GLAST/LAT collaboration towards the 'low-energy' radio bands. Further we will give 
a brief outlook including the extension of this coordinated campaign towards higher 
frequencies and future scientific aims.

\end{abstract}

\maketitle

%%%%%%%%%%%%%%%%%%%%%%%%%%%%%%%%%%%%%%%%%%%%
%% MAINMATTER
%%%%%%%%%%%%%%%%%%%%%%%%%%%%%%%%%%%%%%%%%%%%

\section{Introduction}

The well known and unusual properties of blazar-type AGNs (e.g. extreme variability, 
high degree of polarisation, highly superluminal motions and brightness temperatures 
exceeding the inverse Compton limit; e.g. \cite{Urry1999}) are collectively interpreted 
within the context of emission originating in relativistic jets oriented very
close ($\le\,20-30^{\circ}$) to the line of sight (e.g. \cite{Urry1995}).

The overall emission scenario explaining the double-humped spectral energy distribution 
(SED) of blazars (e.g. synchrotron emission plus synchrotron self-Compton (SSC) 
and/or external Compton (EC), hadronic processes) is still poorly understood. 
So is the case for the origin of the observed flux density and polarisation 
variability on time scales of weeks to months. The study of this rapid 
blazar variability in the radio bands provides insight into the AGN structure 
on linear scales or flux density levels not accessible even to interferometric 
imaging. Here, different models are discussed such as shock-in-jets 
(e.g. \cite{Marscher1996}) or colliding relativistic plasma shells (e.g. 
\cite{Guetta2004}). Further, in the case of precessing binary black-hole 
systems, rotating helical jets or helical trajectories of plasma elements, 
they suggest changes in the direction of forward beaming, thus introducing 
flares due to the lighthouse effect (e.g. \cite{Camenzind1992}, \cite{Begelman1980}, 
\cite{Villata1999}). Hence, variability studies furnish important clues about 
size, structure, physics and dynamics of the emitting region. Here, AGN/blazar 
monitoring programs are of great importance as they provide the necessary 
observational constraints for the different theoretical models towards 
understanding the physical origin of energy production in AGN.

A powerful tool in the study of relativistic jets is the analysis and modeling
of the simultaneous spectral and temporal behavior of blazars, over 
frequency bands as broad as possible (ideally covering the whole SED from 
radio to TeV energies). In particular, it allows the detailed study of different 
emission mechanisms and variability scenarios. So far, this was usually limited 
to particular well suited cases, e.g. BL Lacertae and 3C\,279 (e.g. \cite{Ravasio2003}, 
\cite{Wehrle1998}) often combining not truly simultaneous broad-band data. The 
demand for quasi-simultaneity is important in view of the high activity and 
rapid variability of these sources. Here, the lack of continuous, 
(quasi-) simultaneous observations at all wavebands and the historical lack of 
sufficient $\gamma$-ray data hampered past efforts to understand and study in 
detail the broad-band jet emission. This situation will dramatically change 
thanks to the launch of the GLAST satellite. The LAT detector on-board GLAST 
with a sensitivity of a factor of $\sim$\,30 increased over EGRET is expected to 
detect more than a thousand blazars and to observe $\gamma$-ray spectra resolved 
at a variety of time scales. Furthermore, dense sampled flux monitoring 
data is expected due to LAT's large field-of-view (2.4\,sr) and the GLAST's 
survey observing mode. Consequently, GLAST will provide a tremendous opportunity 
for future, systematic blazar studies addressing important, still open questions 
about the physical processes involved. 

\section{GLAST dedicated multi-wavelengths observations}
In order to fully benefit from these offered GLAST capabilities and to interpret 
the high energy data in comparison with theoretical models, the GeV $\gamma$-ray 
observations have to be combined with an extensive suite of multi-wavelength (MW) 
observations of satisfactory spectral and temporal coverage. Consequently, dedicated 
MW monitoring campaigns are required which will produce comprehensive 
and complementary observations for a large number of potential GLAST blazars 
(see also Tosti et al., current proceedings). 

In this framework, the GLAST/LAT AGN science group is planning extensive broad-band 
campaigns including ToO observations of flaring sources, MW Planned Intensive 
Campaigns (PICs) and MW long-term monitoring (Tosti et al., current proceedings). 
These activities also include proposals to facilities such as Chandra, RXTE, 
Spitzer, Suzaku, {\it INTEGRAL}. In addition, IR\,/\,optical monitoring observations 
are performed with several telescopes.  
 
\section{The 'low-energy' bands: coordinated cm- to mm-monitoring of potential 
GLAST blazars}
In order to complement these MW activities towards the 'low-energy' synchrotron part of 
blazar SEDs, we initiated a GLAST dedicated monitoring campaign in a mutual effort 
between the MPIfR and the Caltech group. Both groups jointly are now starting a 
broad-band flux density and polarization monitoring of potential GLAST blazars.
It is planned that at 15 GHz the Owens Valley Radio Observatory (OVRO) 40m telescope 
will soon start to monitor a large number of radio sources per day (about 1000 GLAST 
candidates), thus providing e.g. a quasi-continuous record of activity (flaring states, 
'variability index', duty cycles etc.) that can later be combined with 
(quasi-) simultaneous MW - in particular GLAST $\gamma$-ray - observations. 

%Furthermore, this campaign will provide by far the most comprehensive study of radio 
%variability ever attempted in AGN when compared to current/past monitoring campaigns, 
%e.g. Aller et al. (2003), Ter\"asranta et al. (1998). 
%Complementary, a sub-sample of initially 50 (later more) sources are now monitored with 
%the Effelsberg 100m RT with broad spectral coverage (2-43 GHz) every 3-4 weeks. 

Complementary, a sub-sample of initially $\sim$\,50 sources (later more) are being 
monitored with the 100m Effelsberg radio telescope (EB). The Effelsberg telescope provides 
high sensitivity, fast frequency-switching and polarisation capabilities, as well as 
a broad frequency coverage. Consequently, EB is the ideal instrument to obtain high-precision, densely 
time-sampled and (quasi-) simultaneous broad-band (11\,cm to 7\,mm) data for a 
larger sub-sample of sources in a reasonable amount of observing time. 
The EB monitoring program started recently (January 2007) with observations 
at 110, 60, 36, 28, 20, 13, 9 and 7\,mm wavelengths which are planned to be continued over the 
next years with a sampling of one epoch every 3--4 weeks. The present source list 
comprises a list of $\sim$\,50 'famous', bright and flat-spectrum blazars selected from 
the 'high priority blazar list' of the LAT AGN science group, i.e. those sources which 
will be of highest interest for later MW studies with the LAT instrument. The sources 
were selected such that (i) the list includes all sources which will be target of future, 
coordinated and intensive MW campaigns by the LAT AGN group, and (ii) a maximum overlap 
with current VLBI projects is achieved, i.e. 35 of the selected objects are also part of 
the MOJAVE monitoring program (see also Lister et al., Kadler et al. current proceedings) 
and thus are regularly observed with VLBI.

%In addition, (quasi-) simultaneous observations are provided by optical facilities. 
%The extension of the Effelsberg/OVRO monitoring program towards short mm-wavelengths 
%is the main target of this proposal. In addition, (quasi-) 

\section{Towards higher frequencies (mm-, sub-mm bands, IR\,/\,optical)}
An important next step is to complement this monitoring efforts with observations 
at shorter wavelengths (mm\,/\,sub-mm). In the standard synchrotron scenario of evolving 
outbursts, the flux density variability appears first at higher frequencies 
(e.g. optical/X-ray) and then propagates though the spectrum towards longer wavelengths. 
Here, the variability at short mm\,/\,sub-mm bands is usually much more pronounced and faster 
than in the longer cm-radio regime. Hence, complementary observations at short 
mm to sub-mm bands are crucial as they provide the important link between the long wavelengths 
radio bands (OVRO/EB) and the more energetic IR\,/\,optical\,/\,X-ray regime, where blazars 
usually show their maximum synchrotron output. 

The IRAM 30\,m telescope on Pico Veleta (PV) is the ideal instrument to effectively fill 
this gap (3, 2, 1.3\,mm) due to its high sensitivity, frequency coverage/agility and now 
also polarisation capabilities. Hence, a coordinated effort between EB/OVRO and PV 
is planned and was proposed to combine and intensify the ongoing monitoring at 
PV (\cite{Ungerechts1998}). If successfull, this coordinated campaign will provide
densely time-sampled, precise and (quasi-) simultaneous broad-band (11cm to 1\,mm) 
spectra and variability data for a large number of potential GLAST blazars. Additionally, 
efforts are currently being put in employing sub-mm facilities as well in order to further 
extend the wavelength coverage towards the sub-mm bands.

In addition, we further aim at extending these broad-band monitoring efforts with quasi-simultaneous 
observations in the IR\,/\,optical regime. Here, observing time was already approved and 
observations (V/R/I/H bands) will be performed with the 1.2m Kryoneri telescope (Greece), 
the Rapid Eye Mount (REM) in Chile and the Perugia automatic telescope in Italy 
(see also Tosti et al., current proceedings).

\section{Outlook}
%- increase the number of sources 
Finally, we will give a short summary of the possibilities which this monitoring campaign 
will provide for future blazar studies in the upcoming GLAST era. The broad-band monitoring 
presented here will in detail allow to perform:

\noindent
{\bf (i) Variability and spectral evolution studies across the cm- to (sub-) mm-bands:}
This includes e.g. the study of the frequency dependent variability of a large blazar 
sample, e.g. to search for correlations, time lags etc. in comparison with variability models; 
the determination of variability Doppler factors and a first-time systematic study of polarisation 
variability at mm-wavelengths in comparison with the cm-regime; the study of simultaneous radio 
spectra (which are not affected by the overall variability) and their evolution (spectral indices, turnover) 
in comparison with synchrotron/variability models (e.g. \cite{Turler2000}, \cite{Lindfors2005}). 
The polarisation information will further allow to e.g. determine magnetic fields and systematically 
study rotation measures. 

\noindent
{\bf (iii) VLBI related studies:}
Here, a strong overlap with VLBI and AGN jet physics exists. Since many sources of our sub-sample are 
included in current VLBI projects (MOJAVE as well as 3\,mm GMVA monitoring), a 
%in particular, we are strongly 
%involved in the 3\,mm GMVA monitoring and 2cm-VLBA survey/MOJAVE project (e.g. Kellermann et al. 
%1998, Zensus et al. 2002). Aspects: 
combination with quasi-simultaneous VLBI data becomes possible. This will allow to directly relate 
the observed VLBI structure and jet kinematics to the overall single-dish flux density, spectral 
variability and Doppler factors, e.g. to study possible flare-ejection relations and to identify the 
jet-regions responsible for the cm\,/\,mm-variability (e.g. \cite{Bach2006}).

% and, in particular, 
%the variability/emission seen by GLAST in the $\gamma$-ray band. 

\noindent
{\bf (iii) Broad-band studies in the GLAST era:}
The monitoring efforts presented here will complement and support the future MW activities 
(IR\,/\,optical\,/\,X-ray\,/\,$\gamma$-ray\,/\,TeV) of the GLAST/LAT collaboration towards the 
'low-energy' radio bands. In this framework, future MW studies will include e.g. (quasi-) 
simultaneous, time-resolved broad-band (radio to $\gamma$-ray-) spectra in comparison 
with jet emission models (leptonic vs. hadronic models, SSC\,/\,EC etc.); the interplay between 
the variability across all bands (correlations, time lags etc.) - in particular in view of the GLAST
$\gamma$-ray data - and the study of geometrical and radiation induced variability. In combination 
with VLBI, these efforts will furthermore allow to investigate the jet regions responsible 
for the $\gamma$-ray emission and to study in more detail e.g. the relation between $\gamma$-ray
flares and newly ejected jet components on VLBI scales.

\end{document}